# Nonlinear optics in Xe-filled hollow-core PCF in high pressure and supercritical regimes

M. Azhar,[1,*] N. Y. Joly,[2,1] J. C. Travers[1] and P. St.J. Russell[1,2]

[1]Max Planck Institute for the Science of Light and [2]Department of Physics, University of Erlangen-Nuremberg, Günther-Scharowsky-Str. 1, 91058 Erlangen, Germany
*Corresponding author: mohiudeen.azhar@mpl.mpg.de

Supercritical Xe at 293 K offers a Kerr nonlinearity that can exceed that of fused silica while being free of Raman scattering. It also has a much higher optical damage threshold and a transparency window that extends from the UV to the infrared. We report the observation of nonlinear phenomena, such as self-phase modulation, in hollow-core photonic crystal fiber filled with supercritical Xe. In the subcritical regime, intermodal four-wave-mixing resulted in the generation of UV light in the $HE_{12}$ mode. The normal dispersion of the fiber at high pressures means that spectral broadening can clearly obtained without influence from soliton effects or material damage. © 2013 Optical Society of America

*OCIS Codes: 190.4370, 060.5295*

The long diffraction-free interaction lengths provided by hollow-core photonic crystal fiber (HC-PCF) make it uniquely versatile for the study of nonlinear optics in gases, offering a high damage threshold and orders of magnitude lower transmission loss compared to capillaries of the same (~20 μm) core diameter [1,2]. A special sub-class of these fibers is kagomé-style HC-PCF, which provides ultra-broadband optical transmission at losses of below 1 dB/m, the guidance mechanism being a form of two-dimensional anti-resonant reflection [3]. In addition, when evacuated, kagomé-PCF has weak and smoothly wavelength-dependent anomalous dispersion that can be balanced against the normal dispersion of a filling gas, permitting the dispersion landscape to be pressure-tuned. In previous work on Ar-filled kagomé-PCF, this enabled observation of tunable deep-UV (DUV) light via dispersive wave generation from self-compressed fs pulses at 800 nm [4] and the first observation of a plasma-driven soliton blue-shift [5-6]. In this paper we report on ultrafast nonlinear dynamics in kagomé-PCF filled with high pressure and supercritical Xe. Previous studies of supercritical gases have included Brillouin scattering in Xe [7] and Raman scattering in $CO_2$ [8].

Using tabulated density data at ambient temperature (293 K) [9], the $n_2$ values for Ar, Kr and Xe are plotted against pressure in Fig. 1(a), assuming that $n_2$ is proportional to density. Note that the critical points are (48 bar, ~150 K) for Ar, (55 bar, ~209 K) for Kr and (58 bar, ~289 K) for Xe. At room temperature the influence of the critical point is weak for Kr and Ar and therefore the gas density, and hence $n_2$, varies more or less linearly with pressure, reaching respectively ~5% and ~23% of the value in fused silica glass at 150 bar [10]. For Xe, the influence of the critical point is much stronger, leading to a sharp increase in $n_2$ when the pressure reaches ~60 bar [9]; recent linear measurements have confirmed this [11]. When the pressure and temperature lie above the critical point – easily achievable in experiment without the need for a cryogenic system [12] – Xe becomes a supercritical fluid, with a nonlinearity that can exceed that of fused silica.

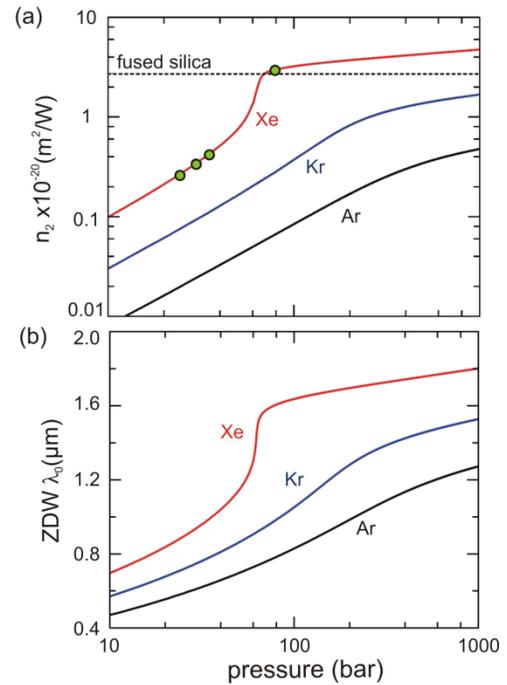

Fig. 1: (a) The dependence of nonlinearity $n_2$ on pressure at 293 K for Xe, Kr and Ar. The nonlinearity of fused silica is included for comparison. For Xe the curve starts out linear and then sharply increases close to the critical point. The green circles mark the pressures at which the experiments were performed. (b) Pressure-dependence of the zero dispersion wavelength in a gas-filled kagomé-PCF with a core diameter of 18 μm. The dispersion is anomalous for wavelengths longer than $\lambda_0$.

Theory predicts that Xe will exhibit a temporally non-local (response times in the μsec range) nonlinearity close to the critical point [13], due to intense scattering arising from critical opalescence. In this paper we avoid this regime by operating sufficiently above or below the critical point, where Xe remains transparent.

As described in previous papers, the dispersion of the guided mode in kagomé-PCF can be described using a capillary model [14,15]. As the pressure increases, the normal dispersion of the gas counteracts the weak anomalous waveguide dispersion of the empty kagomé PCF, creating a pressure-tunable zero dispersion wavelength (ZDW, $\lambda_0$). Fig. 1(b) shows the variation of $\lambda_0$ with pressure for Ar, Kr and Xe. Although Xe clearly extends the range of tunability of $\lambda_0$ compared to Ar and Kr, opalescence makes it unusable in the vicinity of the supercritical transition. Note that the dispersion can also be tuned by changing the fiber core diameter, for example, $\lambda_0$ could be shifted further into the infrared with larger core diameters [2]. At high enough pressure the dispersion becomes normal for all wavelengths of interest, enabling the observation of effects such as intermodal four-wave mixing (iFWM) without interference from soliton dynamics, which usually dominates in the anomalous dispersion region.

The experimental set-up consisted of a 28 cm length of kagomé-PCF (core diameter 18 µm) with a high pressure gas cell at each end. The pump laser was an amplified Ti:sapphire system (wavelength 800 nm) delivering pulses of duration 150 fs and energy ~1.8 µJ at a repetition rate of 250 kHz. Diagnostics included a UV-sensitive camera for modal imaging and a spectrometer sensitive from 200 to 1100 nm. To prevent the spectrometer from saturating, the signal was attenuated by reflection at two wedged glass plates. A parabolic mirror was then used to focus light into the spectrometer.

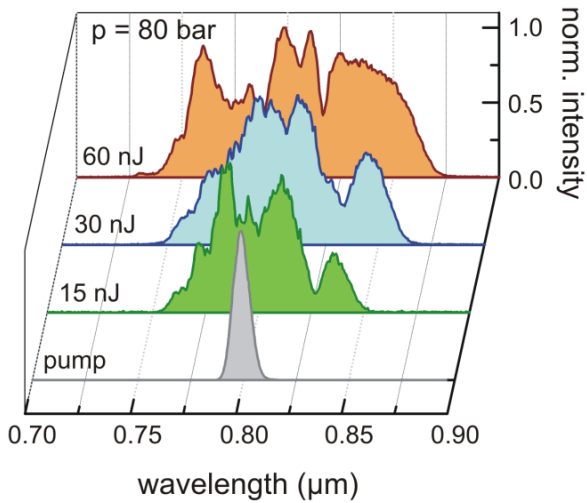

Fig. 2: Experimental output spectra (a) spectral broadening due to SPM in supercritical Xe at 80 bar for pulse energies in the fiber 15, 30 and 60 nJ

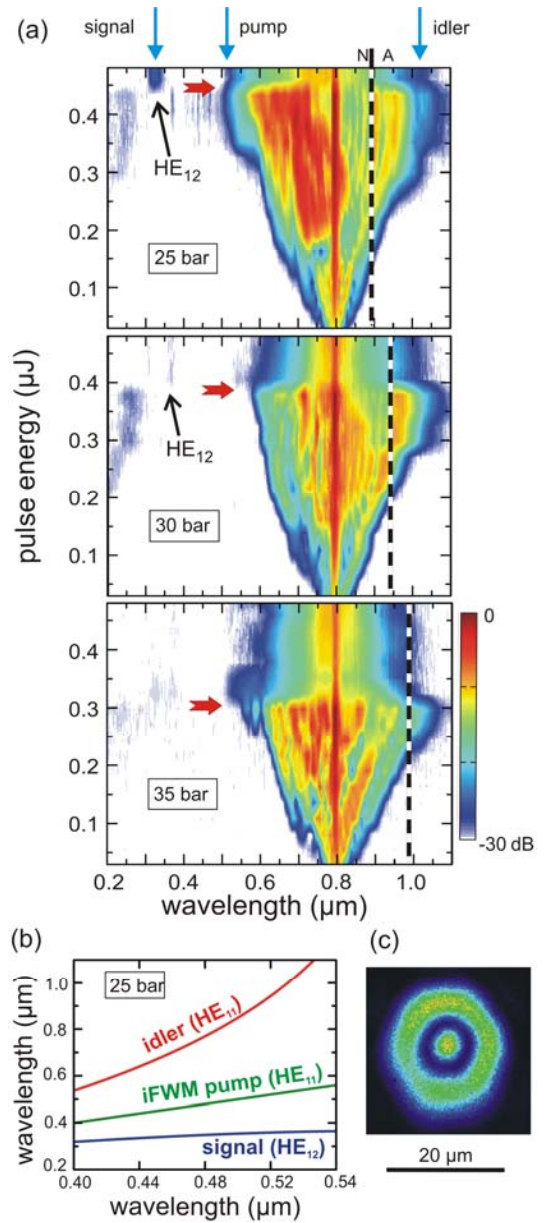

Fig. 3: (a) Experimental spectral broadening with launched pulse energy at (left to right) 25 bar, 30 bar and 35 bar; the dashed vertical lines indicate the position of $\lambda_0$ (N = normal, A = anomalous); the red arrows indicate the onset of self-focusing in the input cell. (b) Theoretical phase-matching wavelengths ($2/\lambda_P = 1/\lambda_S + 1/\lambda_I$) for iFWM at 25 bar; for a 510 nm pump, the signal and idler wavelengths are 325 and 1184 nm. (c) Experimental near-field image of the light emitted in the HE$_{12}$ mode at 325 nm (indicated by the black arrow in (a)).

Supercritical Xe was collected by liquefying Xe in steel pipes cooled by dry ice. After a sufficient amount of Xe had collected, the pipes were warmed up to room temperature. This simple procedure allowed us to reach Xe pressures of 200 bar from a 40 bar gas cylinder, while maintaining high Xe purity.

We filled the fiber with 80 bar Xe, well inside the supercritical regime at 293 K. At this pressure the nonlinear refractive index is ~$2.8\times10^{-20}$ m$^2$/W [16, 17], which matches the value for fused silica [18] (Fig. 1(a)). As the launched pulse energy was increased, self-phase modulation (SPM) caused spectral broadening (Fig. 2). This continued up to ~80 nJ, when the broadening abruptly collapsed, a dramatic effect that we attribute to disruption of the in-coupling by self-focusing effects in the input gas-cell. To verify this, we performed experiments in a simple gas cell, obtaining reasonable agreement with full spatio-temporal numerical simulations using the methods described in [19], and simple numerical estimates of nonlinear focusing described in [20].

The disruptive effects of self-focusing are also apparent in the sub-critical regime (Fig. 3(a)), where spectral broadening is abruptly attenuated above a certain critical launched energy (marked by the red arrows in Fig. 3(a)) that depends inversely on the pressure. This spectral collapse is accompanied by a 70% drop in transmitted power. At 25 bar, $\lambda_0 \sim$ 890 nm and the pump wavelength lies in the normal dispersion regime. In addition to SPM-induced spectral broadening, an unexpected band of UV light appears at ~325 nm. Using a narrow-band filter to isolate the near-field pattern at this wavelength (Fig. 3(c)), we were able to identify this signal as being in the HE$_{12}$ mode. We attribute its appearance to intermodal four-wave mixing (iFWM). Fig. 3(b) shows the results of a phase-matching analysis, based on the Marcatili model [14], assuming that pump and idler are in the HE$_{11}$ mode and signal in the HE$_{12}$ mode. As the spectrum broadens, it reaches beyond 1 μm wavelength and is then able to act as an HE$_{11}$ idler seed for iFWM, pumped by the green spectral edge at 510 nm. These two signals result in the generation, via iFWM, of signal photons in the HE$_{12}$ mode at 325 nm. The analytical theory predicts wavelengths that are in good agreement with the observations; the slight disagreement can be attributed to deviations of the actual fiber dispersion curve from that predicted by the Marcatili model. The weak iFWM signal at ~375 nm and 30 bar, visible in the middle panel of Fig. 3(a), was also experimentally confirmed to be in the HE$_{12}$ mode.

In conclusion, clear SPM broadening is observed in a HC-PCF filled with supercritical Xe, which at 80 bar has the same Kerr nonlinearity as silica. As a result of self-focusing effects in the launching cell, the spectral broadening was observed to collapse abruptly at a critical energy level that scaled inversely with the gas pressure; this effect could be eliminated by placing the glass window closer to the fiber end-face. Intermodal four-wave mixing was observed to occur at the point of spectral collapse, resulting in generation of UV light in the HE$_{12}$ mode. Compared to all-silica fiber systems, noble-gas-filled HC-PCF offers a much higher damage threshold, excellent transparency at ultraviolet wavelengths, pressure-tunable dispersion and Raman-free operation. The result is a remarkably flexible system for nonlinear fiber optics.